**Radiation of Channeled Positrons in Ultrasonic Wave**

E.A. Ayryan [1] , K.V. Ivanyan [2]* , M. Hnatic[3,4,5] ,
P. Kopcansky [3] , K.B. Oganesyan [6] , V.V, Papoyan [1,4] , M. Timko [3]

[1] LIT, Joint Institute for Nuclear Research, Dubna, Russia

[2] M.V. Lomonosov Moscow State University, Moscow 119991, Russia

[3] Institute of Experimental Physics SAS, Kosice, Slovakia

[4] BLTP, Joint Institute for Nuclear Research, Dubna, Russia

[5] Faculty of Sciences, P. J. Safarik University, Kosice, Slovakia

[6] A. Alikhanyan National Lab, Yerevan Physics Institute, Yerevan, Armenia

In the classical approximation, we examine the problem of radiation by channeled positrons in the field of a longitudinal ultrasonic wave when the condition of parametric resonance is satisfied. We show that in the case of planar channeling the intensity of the spectral distribution of the radiation can be several times larger than when the resonance condition is not satisfied.

## 1. Introduction

Usually in the calculation of the radiation from a particle in planar channeling one uses the averaged potential of the atomic planes. Under certain conditions, however, the traveling particle may "feel" the separation $d_s$ between the atomic chains forming the planes. These conditions can be realized, in particular, at large amplitudes of transverse oscillations of the particle in situations approaching dechanneling. Then, the approximation of constant potentials on the atomic planes is inadequate and it is necessary to allow for the influence of discrete atomic chains. This allowance was made in [1] , the case of macroscopic channeling by expanding the real potentials of the atomic in Fourier series and considering the motion of the particle in such a potential as classical.

-----------------

*k.ivanyan@yandex.com

In particular, for the case when the atomic chains forming the planar channel lie in both planes without shifting, the Mathieu equation was obtained for the motion in the xz plane, transverse to the direction of the entry of the particle into the crystal (z axis); this describes the so-called resonance dechanneling of particles, which arises under certain conditions. There are also experiments confirming the possibility of resonance dechanneling.

The classical theory of radiation by channeled particles in presence of an external field periodically perturbing the crystal was first given in [2]. The quantum theory of the effect is expanded in [3,4 ] .

We note that if the angle of planar channeling is much larger than the Bragg angle, as is well known, we can give a classical description of the motion of the particle in the channel, which besides its relative simplicity, allows an analogy with well-studied cases of parametric resonance.

In [5,6**],** in further development of [1], the classical approximation is used to study radiation from an ultrarelativistic positron entering a planar channel of a crystal with initial velocity $v_0 \approx c$ at an angle $\vartheta_0$ less than the critical angle for scattering, in the presence of an external sound wave. It is supposed that the sound wave (hypersound or ultrasound) has the form of a standing wave in the direction of the z axis, along which the channeled particle enters the crystal. In the case where the sound wave is longitudinal, there is a new periodic structure with period equal to the sound wavelength $\lambda_s$ which is greater than the separation between atomic chains, $\lambda_s > d_s$. Since the wavelength can be varied, the realization of this resonance condition is facilitated. Macroscopic channeling of electrons in relativistic strophotron type free electron laser and microscopic channeling of positrons in ionic crystals are discussed in [7-42].

## 2. Equations of Motion

Assuming that the external sound wave weakly perturbs the uniform potential of the atomic planes, the potential energy of a positron in the channel in the field of the external longitudinal sound wave can be described, in the harmonic approximation and in the first order of the displacement of the atomic lattice in the field of the wave, in the following form [1,5,6]

$$U(x,z) = A\cos(2\pi z/\lambda_s) + V_0[1-\mu\cos(2\pi z/\lambda_s)]x^2. \qquad (1)$$

Here x is measured from the median plane, $V_0 = 4U_0/d^2$, $U_0$ is the maximum value of the potential energy of the positron in the channel, d is the width of the channel, ***A*** is a constant; $\mu$ is a small parameter related to the power *I,* of the sound wave: $I_s \approx \frac{1}{2}\rho v_s(\mu\omega_s d_s)$; $\rho$ is the matter density in g/cm3, and $v_s$ and $\omega_s$ are the velocity and frequency of the sound wave. In (1) the time dependence of the potential is neglected; this is valid provided that $\omega_s\tau << 1$, where $\tau$ is the time of flight of the channeled positron across the crystal.

Since the angle $\vartheta_0$ of the entry of the particle into the crystal is small enough ($\vartheta_0 << 1$ ), the transverse velocity of the particle in the planar channel satisfies the condition $v_\perp \approx v_0\vartheta_0 << v_z \approx v_0$ . Taking account of this condition, we can write the equation of motion of the ultrarelativistic positron in the potential (1) in the form

$$\frac{d^2x}{dt^2} + \omega_0^2\left(1-\mu\cos\frac{2\pi z}{\lambda_s}\right)x = 0, \qquad (2)$$

$$\frac{d^2z}{dt^2} = -\frac{1}{c^2}\frac{dx}{dt}\frac{d^2x}{dt^2}\frac{dz}{dt}, \qquad (3)$$

where $\omega_0 = (2V_0/m_0\gamma)^{1/2}$, $m_0$ is the rest mass of the positron, $\gamma = (1-\beta^2)^{-1/2}$ , $\beta = v_0/c$. Replacing z in (2) by $\bar{v}_z t$, where $\bar{v}_z$ is the velocity of the particle averaged over the period of the crystal, we obtain for the motion of the particle in the transverse direction the Mathieu equation

$$\frac{d^2x}{dt^2} + \omega_0^2\left(1-\mu\cos\Omega t\right)x = 0, \qquad (4)$$

where $\Omega = 2\pi\bar{v}_z/\lambda_s$. Assuming that the parameter $\mu$ is small ( << 1 ), it is possible even in the first approximation to observe [1,5,6] in the oscillatory system described by (4) a parametric resonance at $\omega_0 = \Omega/2$, which has very great practical interest. For a particle with energy 5 GeV and $U_0$ = 22.8 eV, d = 1.26 A, and $\omega_0 = 4.5\times10^{14}s^{-1}$ the sound wavelength determined by the resonance condition $\omega_0 = \Omega/2$ is 2.4 A, which corresponds to a frequency $\omega_s \approx 10^9 s^{-1}$. The transit time $\tau << 1/\omega_s = 10^{-9}s$ ; meaning apparently that a crystal of thickness $l = c/\omega_s << 30cm$ , that is, no more than a few centimeters, can be used. We shall see below, however, that the presence of other, more stringent conditions significantly reduces the allowable thickness of the crystal.

In [5,6], with the assumption $\mu << 1$, Eq. (4) was solved by the method of successive approximations. In the first nonvanishing approximation they found the trajectory and speed of the particle, and the maximum frequency of the radiation and its intensity for the first few harmonics. The results obtained [5,6] however, are valid far from resonance, when $\Omega/2$ is not too close to $\omega_0$: $|\Omega/2-\omega_0| >> \mu\omega_0$ and the expression for the intensity of radiation is not fundamentally different from the corresponding expression obtained in [42] with the assumption of a harmonic potential for the atomic planes, $U(x)=V_0x^2$. It is interesting to consider a solution of Eq. (4) in the immediate neighborhood of resonance, $\omega_0 \approx \Omega/2$. The amplitude of the transverse oscillations of the trajectory of the particle grows in this case, which may lead under specific conditions to a more notable increase in the intensity of the radiation by the channeled particle in the resonance regime.

Assuming that $\mu << 1$, we can consider for Eq. (4) in first-order approximation only the resonance appearing at $\omega_0 = \Omega/2$, and construct an approximate solution of this equation by following the asymptotic method of Bogolyubov and Mitropolskii [43]. The solution of Eq. (4), satisfying the condition x = 0 at t = 0 can be written in the form

$$x(t) \approx x_m e^{\lambda t} \sin(\Omega t/2), \qquad v_x(t) \approx 2^{-1/2} v_\perp e^{\lambda t} \cos(\Omega t/2). \tag{5}$$

where

$$x_m = v_\perp / 2^{1/2}\omega_0(1+\mu/4) \approx v_\perp / 2^{1/2}\omega_0.$$

$v_\perp \approx v_0\vartheta_0$ is the initial transverse velocity of the positron, and $\lambda$ is real and positive and is determined by the following expression

$$\lambda = \left[\mu^2\omega_0^4/4\Omega^2 - \left(\omega_0-\Omega/2\right)^2\right]^{1/2} \approx \mu\omega_0/4. \tag{6}$$

The condition of reality of $\lambda$ is to first order in $\mu$

$$\mu\omega_0/4 > |\omega_0-\Omega/2|. \tag{7}$$

Thus, if the frequency $\Omega$ of the external perturbation is in the interval

$$2\omega_0(1-\mu/4) < \Omega < 2\omega_0(1+\mu/4), \tag{8}$$

then in the system there appears the principal parametric (demultiplicative) resonance, at which the amplitude of the transverse vibrations grows exponentially in time r. The inequality (8) determines the region of instability inside which the equilibrium position x = 0 is unstable and oscillations arise spontaneously in the system, and lead under certain conditions to resonance dechanneling. The longitudinal component of velocity u, is determined by the condition that the total velocity $v_x^2 + v_z^2 = v_0^2$ be constant under the condition $v_x^2 << v_0^2 \sim c^2$, and is equal to

$$v_z = v_0\left(1 - 1/2 v_x^2 / v_0^2\right). \qquad (9)$$

From the requirement $v_x^2 / v_0^2 << 1$ it follows that $0 \le \lambda t \le 1$. We note that if $\lambda t >> 1$, the formulas obtained are inapplicable; at such flight times dechanneling of the particles takes place. The time $\tau$ of the flight of the particle across the crystal, in the course of which we can use the formulas obtained above, is bounded by the conditions $\tau << 1/\lambda = 4/\mu\omega_0$, and $\omega_s \tau << 1$. The resulting condition $4/\mu\omega_0 << 1/\omega_s$, means $v_s / c << \mu / 8 << 1$, that is, the intensity of the sound wave must be greater than the limit determined above. The velocity of sound in solids (quartz) is $v_s \approx 6\times10^5$ cm/s, that is $v_s / c \approx 2\times10^{-4}$ and consequently $\mu >> 10^{-3}$. We pick $\mu = 0.1$. Then for a positron with energy $E_0 \sim$ 5 GeV and for $\omega_0 =$ 22.8 eV, d = 1.26 A, we have $\omega_0 \approx 4.5\times10^{14} s^{-1}$, the thickness of the crystal $l \le c\tau = 4c/\mu\omega_0 = 3\times10^{-3}$ cm, and the power of the ultrasound wave $I_s \approx 0.12\omega / cm^2$.

The period T of the oscillation of the particle in the crystal is $2\pi/\omega_0 = 4\pi/\Omega$ and the distance traveled in one period is $l_0 = cT = 2\pi c/\omega_0 = 4\pi c/\Omega$; consequently $l = l_0(2/\pi\mu)$ and in its entire passage through the crystal the particle succeeds in making a great number of oscillations; during this time $\lambda t$ can grow to a magnitude of about 1. Thus, for the whole time of flight across the crystal, the factor $e^{\lambda t} t$ slowly and monotonically changes from 1 to **3** and on averaging over time it is possible to neglect the change in this factor, replacing $\exp(\lambda t)$ in the formulas (5) by $\exp(\lambda t^*)$, where ***t*** * is an effective time of fight of the particle across the crystal. For this case, we have

$$\begin{aligned} &x(t) = x_m e^{\lambda t} \sin\frac{\Omega}{2}t, \qquad v_x(t) = \frac{v_\perp}{\sqrt{2}} 2e^{\lambda t} \cos\frac{\Omega}{2}t, \\ &v_z(t) = \overline{v}_z - \delta v_z \end{aligned} \qquad (10)$$

Where

$$\overline{v}_z = v_0 \left[ 1 - \frac{1}{4} \left( \frac{x_0 \omega_0}{c} \right)^2 \right],$$

and $\delta v_z$ is the oscillatory part of the longitudinal velocity, equal to

$$\delta v_z = \frac{1}{4} v_0 \left( \frac{x_0 \omega_0}{c} \right)^2 \cos 2\omega_0 t, \quad x_0 = x_m e^{\lambda t^*}.$$

It is easy to verify that the solutions (5) and (10) satisfy Eq. (3) with accuracy to higher order than $(x_0 \omega_0 / v_0)^2$.

### 3. The Intensity of Radiation

For the calculation of the intensity of radiation when the condition $\omega_0 \approx \Omega / 2$ is fulfilled, we use formulas obtained in [44] for the spectral and angular distribution of the intensity of the $k$ - th harmonic of the radiation, replacing $x_m$ these formulas with $x_0 = x_m \exp(\lambda t^*)$ ($x_m$ is the initial amplitude corresponding to the harmonic potential). In the dipole approximation, when the condition $x_0 \omega_0 / \mathrm{c} << 1/\gamma$ is satisfied, it radiates mainly the first harmonic k = 1, the maximum frequency of the radiation is $\omega_m \approx 2\gamma^2 \omega_0$ and hardly differs from that in the absence of resonance. For the spectral distribution of the radiation in the dipole approximation *(k* = 1 ) we have

$$\frac{dI^{res}}{d\omega} = \frac{dI^{nonres}}{d\omega} e^{2\lambda t^*}. \tag{11}$$

For $\lambda t^* \leq 1$, we have $\exp(2\lambda t^*) \leq 10$ and the intensity of the radiation increases by several times. We note that the analogous result for a sound wave transverse to the direction of motion was obtained in Ref. 4 (neglecting absorption). The use of the classical approach is limited by the condition that the energy of the radiated photons be small compared to the energy of the positron: $\hbar\omega / mc^2 \gamma << 1$. On fulfilment of the condition for the dipole character, this limit leads to the following inequality: $\lambda_s >> 2\pi \lambda_c \gamma$, where $\lambda_s$ is the wavelength of the ultrasound, and $\lambda_c = \hbar / mc = 2.46 \times 10^{-10} cm$ is the Compton wavelength of the positron. As is clear from this inequality, for $\lambda_s = 2.4 \times 10^{-4} cm$ the value of $y$ can vary in a fairly wide range: from 10 to $10^4$. If the dipole condition is not satisfied, $1/\gamma << x_0 \omega_0 / c << 1$, the maximum radiation frequency of the harmonics in the case considered is

$$\omega_{km} \approx 4\omega_0 k (x_0 \omega_0 / c)^{-2}$$

and are decreased by a factor $\exp(2\lambda t^*)$ compared to the nonresonant case. The condition for the validity of the classical method is

$$\lambda_s >> 4\pi\lambda_C \gamma k (x_0 \omega_0 \gamma / c)^{-2}$$

In the relativistic case ($\gamma >> 1$) and in the range of radiation angles $\vartheta << 1$, where the bulk of the radiated intensity is concentrated, the intensity of the radiation in the different harmonics $k \geq 1$ can be represented in the form"

$$\frac{dI_k}{d\xi} = \frac{4e^2\omega_0^2\gamma^2}{c} F(\xi), \quad \frac{dI_k}{do} = \frac{4e^2\omega_0^2\gamma^4}{c} \frac{\xi^2}{\pi k} f(\xi, \varphi), \tag{12}$$

where parameters $F(\xi)$, $f(\xi,\varphi)$ and $\xi$ are presented in [46].

## 3. Summary

From these formulas it is apparent that when the angle of radiation $\vartheta$ is less than the angle of entry of the particle into the crystal: $\vartheta << x_0\omega_0 / c$, and the dipole condition is violated: $1/\gamma \leq x_0\omega_0 / c$, only the odd harmonics are radiated, the angular distribution of intensity $dI_k / do$ decreases by roughly $\exp(-4\lambda t^*)$, and the spectral distribution of intensity of radiation $dI_k / d\omega$ is practically unchanged.

In the general case ($\vartheta << x_0\omega_0 / c$), the dependence of the intensity of radiation on the amplitude $x_0$ of oscillation is quite complicated and a numerical estimate of the spectral and angular dependence of the radiation is necessary. From the results of numerical calculations of the spectral distributions for the first ten harmonics for different values of the amplitude $x_0$ [45] we can assert that the spectral intensity of the radiation is increased several times with growth of the amplitude of the transverse oscillation x, (upon the satisfaction of the conditions indicated above); at the same time the share of the radiation in the higher harmonics increases markedly. We recall, in particular, that the formulas obtained above are valid for crystals of thickness $l \leq 4c / \mu\omega_0$, which in the example we considered comes out as we see to $\approx 3\times10^{-3} cm$. Analogous considerations can be made also for the case of axial dechanneling of electrons.

**Acknowledgements**

KBO and EAA are grateful for the support from the NSP of Slovak Republic.